\documentclass[sigconf,natbib=true]{acmart}

\usepackage{booktabs}
\usepackage{caption}
\usepackage{multirow} 
\usepackage{arydshln} 
\usepackage{booktabs} 
\usepackage{amsmath}
\usepackage{makecell}
\usepackage{soul, color, xcolor}
\usepackage[table]{xcolor}
\usepackage{enumitem}
\usepackage[dvipsnames]{xcolor}
\usepackage[normalem]{ulem}
\usepackage{subcaption}

\usepackage{booktabs}
\usepackage{multirow}
\usepackage{graphicx}

\usepackage{placeins}
\usepackage{changepage}

\definecolor{bgray}{HTML}{F0F0F0}
\definecolor{bblue}{HTML}{E6F0FF}
\definecolor{bgreen}{HTML}{E8F5E9}
\definecolor{borange}{HTML}{FFF3E0}
\definecolor{HeadBG}{gray}{0.95} 
\definecolor{RowAlt}{gray}{0.98} 

\renewcommand\footnotetextcopyrightpermission[1]{}
\AtBeginDocument{%
  }

\begin{document}

\title{Beyond Item Order: Temporal Gap Tokenization for Generative Recommendation with Semantic IDs}

\author{Chengkai Huang}
\authornote{Corresponding author.}
\affiliation{%
  \institution{University of New South Wales,}
  \institution{Macquarie University}
  \city{Sydney}
  \country{Australia}
}
\email{chengkai.huang1@unsw.edu.au}

\author{Tianqi Gao}
\affiliation{%
  \institution{Independent Researcher}
  \city{Hangzhou}
  \country{China}
}
\email{tianqig358@gmail.com}

\author{Hongtao Huang}
\affiliation{%
  \institution{University of New South Wales}
  \city{Sydney}
  \country{Australia}
}
\email{frickhuang@foxmail.com}

\author{Quan Z.	Sheng}
\affiliation{%
  \institution{Macquarie University}
  \city{Sydney}
  \country{Australia}
}
\email{michael.sheng@mq.edu.au}

\author{Lina Yao}
\affiliation{%
  \institution{CSIRO’s Data61}
  \institution{University of New South Wales}
  \city{Sydney}
  \country{Australia}
}
\email{lina.yao@unsw.edu.au}



\renewcommand{\shortauthors}{Chengkai et al.}

\begin{abstract}
Semantic-ID-based generative recommendation has recently emerged as a scalable paradigm for sequential recommendation, where each item is represented by a compact sequence of discrete codes and next-item prediction is formulated as code generation. Existing methods, however, typically construct user histories as sequences of static item identifiers, leaving the elapsed time between consecutive interactions outside the generative input. This temporal blindness is problematic because inter-interaction gaps provide useful cues about interest continuity and preference drift. In this paper, we propose \textbf{ChronoSID}, a lightweight temporal augmentation framework for semantic-ID-based generative recommendation. ChronoSID injects temporal signals into the standard three-stage semantic-ID pipeline from two complementary perspectives. First, we introduce \emph{Time-Aware Field-Aware Masked Auto-Encoding} (TA-FAMAE), which regularizes item representation learning with an auxiliary time-gap prediction objective. Second, we discretize historical interaction intervals into fixed log-scale gap tokens and interleave them with semantic ID tuples as the encoder input of the sequence-to-sequence generator. This design preserves the compact SID generation paradigm while enabling the model to capture time-aware transition patterns. Experiments on Amazon review benchmarks show that ChronoSID consistently improves over ReSID and other competitive generative recommendation baselines. Ablation studies further verify the contribution of both temporal components, and diagnostic analyses show clearer gains under long-gap scenarios where user interests are more likely to drift.
\end{abstract}



\maketitle

\section{Introduction}\label{sec:intro}

Sequential recommendation aims to predict a user's next interaction from her historical behaviors and has long been a central problem in recommender systems~\cite{huang2023dual,huang2023modeling,huang2025towards,ye2026gaussian,ye2025beyond,huang2026dual}. 
Recent advances in \emph{generative recommendation} reformulate this task as an autoregressive generation problem: instead of retrieving the next item by scoring a large candidate set, the model directly generates an identifier of the target item. 
To make this paradigm scalable, a growing line of work represents each item as a compact sequence of discrete \emph{semantic IDs} (SIDs), so that next-item prediction becomes the generation of a short code sequence rather than classification over the entire item vocabulary~\cite{tiger,grhandbook,rpg}. 
Such SID-based recommenders have shown strong potential for large-scale recommendation, as they reduce the output space, enable sequence-to-sequence modeling over structured item codes, and provide a flexible interface between item tokenization and generative modeling.

\begin{figure}
    \centering
    \includegraphics[width=\linewidth]{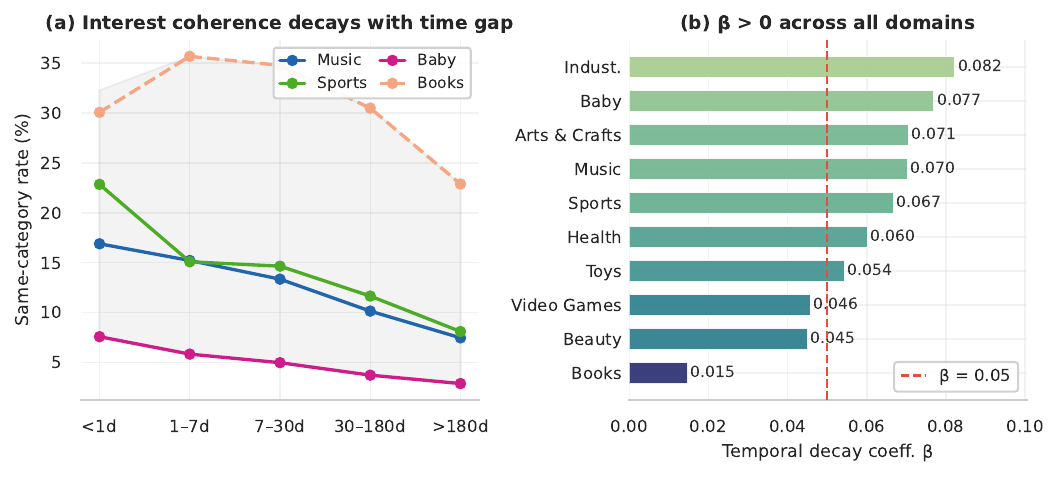}
    \caption{
    Inter-interaction time gaps provide useful cues about interest continuity. 
    Panel (a) shows that the same-category purchase rate generally decreases as the time gap between consecutive interactions increases. 
    Panel (b) reports the fitted temporal decay coefficient $\beta$ across product domains, suggesting that this decay pattern is broadly observed. 
    Since existing SID-based generative recommenders mainly operate on static item-code sequences, such temporal information is not explicitly represented during sequence construction.
    }
    \label{fig:Intro}
\end{figure}

Despite these advantages, existing SID-based generative recommenders typically construct user histories as sequences of static item identifiers. 
Representative methods such as TIGER, LETTER, EAGER, UNGER, and ETEGRec first derive item identifiers from semantic or collaborative item representations and then train autoregressive models over the resulting SID sequences~\cite{tiger,letter,eager,unger,etegrec}. 
Recent methods further improve SID quality through collaborative semantic adaptation, contrastive quantization, tree-structured identifiers, dense-sparse hybrid modeling, and recommendation-native tokenization~\cite{lcrec,cost,seater,liger,cobra,resid}. 
However, once item identifiers are constructed, the interaction history is still usually represented as an ordered sequence of item codes, while the elapsed time between consecutive interactions is left outside the generative input. 
This creates a form of \emph{temporal blindness}: two histories with identical item sequences but very different interaction intervals are encoded in the same way, even though they may correspond to different behavioral states.

Temporal gaps are informative because user interests evolve at different time scales. 
When two interactions occur within a short time window, the user is more likely to remain in the same consumption context; when the user returns after a long interval, her next interaction is more likely to reflect preference drift or a new intent. 
To illustrate this effect, Figure~\ref{fig:Intro} analyzes consecutive interactions in Amazon product domains using the same-category purchase rate as a proxy for interest continuity. 
The rate decreases as the time gap grows, dropping substantially from short-gap interactions to long-gap interactions. 
We further fit a simple temporal decay model for each domain and observe positive decay coefficients in Figure~\ref{fig:Intro}(b). 
These observations do not by themselves solve next-item prediction, but they suggest that time gaps carry behavioral information that is systematically discarded when user histories are reduced to static SID sequences.

This observation is aligned with prior studies on time-aware sequential recommendation, where temporal intervals have been used to model preference drift, recency effects, and time-dependent transitions. 
However, most existing time-aware methods inject temporal information into conventional item-ID recommenders or retrieval-based architectures. 
In contrast, the role of time remains under-explored in SID-based generative recommendation, where both the inputs and outputs are discrete semantic code sequences. 
The key question studied in this paper is therefore:

\begin{adjustwidth}{0.05cm}{0.05cm}
\vspace{1.5em}
\begin{quote}
 \textbf{\textit{How can inter-interaction time gaps be incorporated into SID-based generative recommendation without sacrificing the compactness and scalability of SID generation?}}
\end{quote}
\vspace{1.5em}
\end{adjustwidth}

To answer this question, we propose \textbf{ChronoSID}, a lightweight temporal augmentation framework for SID-based generative recommendation. 
ChronoSID follows the standard three-stage SID pipeline and introduces temporal modeling at two complementary levels, while keeping the quantization stage unchanged for a fair comparison with ReSID~\cite{resid}. 
At the representation level, we introduce \emph{Time-Aware Field-Aware Masked Auto-Encoding} (TA-FAMAE), which augments masked feature reconstruction with an auxiliary time-gap prediction objective. 
This objective regularizes item representation learning with temporal predictability while preserving the original feature-aware reconstruction task. 
At the sequence modeling level, ChronoSID discretizes historical inter-interaction intervals into fixed log-scale \emph{gap tokens} and interleaves them with item SID tuples as the encoder input of the sequence-to-sequence generator. 
In this way, elapsed time is represented as a symbolic input signal alongside semantic item codes, allowing the generator to learn time-aware transition patterns with only a small increase in encoder length.

Our contributions are summarized as follows:
\begin{itemize}[leftmargin=*]
    \item We identify \emph{temporal blindness} as an overlooked limitation of SID-based generative recommendation: existing models mainly generate from static item-code sequences and do not explicitly represent the elapsed time between interactions.
    \item We propose \textbf{ChronoSID}, a lightweight temporal extension of the SID-based generative recommendation pipeline. Our method combines a temporal auxiliary objective for item representation learning with fixed log-scale gap-token interleaving for sequence-to-sequence generation.
    \item We conduct experiments on Amazon recommendation benchmarks and show that ChronoSID improves over strong SID-based generative baselines, including ReSID. Ablation and diagnostic analyses further show that gap-token injection is the main source of improvement and that temporal modeling is particularly useful under long-gap scenarios.
\end{itemize}

\section{Related Work}

\subsection{Sequential Recommendation} 
Sequential recommendation aims to model users' evolving preferences from historical interaction sequences and predict future items \cite{huang2025towards,huang2024foundation}. 
Early neural methods adopt recurrent, convolutional, or graph-based architectures to encode behavior sequences~\cite{narm,caser,gru4rec,srgnn,improvedrnns}. 
For example, GRU4Rec~\cite{gru4rec} models session dynamics with recurrent neural networks, Caser~\cite{caser} captures sequential patterns through convolutional filters, and SR-GNN~\cite{srgnn} represents session transitions with graph neural networks. 
More recently, Transformer-based architectures~\cite{transformer} have become dominant in sequential recommendation. 
SASRec~\cite{sasrec} uses unidirectional self-attention to model item dependencies in historical sequences, while BERT4Rec~\cite{bert4rec} adopts bidirectional self-attention and masked item prediction to learn contextualized sequence representations. 
S$^3$-Rec~\cite{s3rec} further improves representation learning by introducing multiple self-supervised pre-training objectives.
Most conventional sequential recommenders follow a retrieval-oriented paradigm. 
They assign each item a dedicated embedding, encode the user history into a sequence representation, and rank candidate items by matching the sequence representation with item embeddings, often using dot-product or cosine similarity. 
Although approximate nearest neighbor search~\cite{faiss} can improve retrieval efficiency, such methods still rely on large item embedding tables and similarity search over the item space. 
In addition, many sequential models primarily represent user histories as ordered item sequences. 
Temporal information, such as time intervals, recency, and periodicity, has also been studied in time-aware sequential recommendation, where temporal signals are injected into recurrent states, attention scores, or sequence encoders to model preference drift and time-dependent transitions \cite{li2020time,huang2023modeling,huang2023dual}. 
Our work is motivated by the same observation that elapsed time carries useful behavioral information, but focuses on a different setting: semantic-ID-based generative recommendation, where user histories and prediction targets are represented as discrete code sequences rather than item embeddings.

\subsection{Generative Recommendation} 
Generative recommendation formulates next-item prediction as a sequence generation problem, where the model generates an identifier of the target item instead of scoring all candidate items \cite{huang2026generative,huang2026listwise,gao2026factorized}. 
A key challenge in this paradigm is how to define item identifiers that are compact, semantically meaningful, and easy for the generator to predict. 
Semantic ID-based methods address this challenge by representing each item as a short sequence of discrete codes, thereby reducing the output space and enabling sequence-to-sequence recommendation over item identifiers~\cite{tiger,rpg,grhandbook}.
TIGER~\cite{tiger} is a representative early work in this direction. 
It first derives item representations from textual information, quantizes them into discrete semantic IDs using RQ-VAE, and then trains an encoder-decoder model to generate the target item's identifier conditioned on historical interactions. 
During inference, candidate identifiers are produced via beam search and mapped back to items for Top-$K$ recommendation. 
Following this pipeline, subsequent studies have improved different components of SID-based generative recommendation. 
CoST~\cite{cost} enhances semantic tokenization by preserving neighborhood structures during quantization. 
LIGER~\cite{liger} and COBRA~\cite{cobra} combine discrete semantic IDs with dense representations, allowing the generator to exploit both compact identifiers and richer item information. 
Other methods incorporate collaborative signals into identifier learning. 
LETTER~\cite{letter} introduces learnable item tokenization with collaborative supervision, EAGER~\cite{eager} separately models semantic and behavioral identifiers, and UNGER~\cite{unger} learns unified item codes by integrating semantic and collaborative information. 
ETEGRec~\cite{etegrec} further couples item tokenization with downstream generative recommendation in an end-to-end manner.

More recently, ReSID~\cite{resid} proposes a recommendation-native semantic ID framework that aligns item representation learning and semantic quantization under a unified objective. 
Compared with earlier methods that rely heavily on language-derived representations or separately learned tokenizers, ReSID provides a stable three-stage pipeline based on structured recommendation features and globally aligned quantization, achieving strong empirical performance. 
However, these SID-based generative recommenders mainly focus on constructing better static item identifiers. 
After the identifiers are obtained, user histories are still represented primarily as sequences of item-code tuples, while the elapsed time between interactions is not explicitly encoded in the generative input. 
ChronoSID builds on this line of work by preserving the compact SID generation paradigm and augmenting the input sequence with discretized temporal gap tokens, allowing temporal information to participate directly in semantic-ID-based sequence generation.

\section{Problem Formulation}
\label{sec:problem}

Let $\mathcal{V}$ denote the item set. 
For each user $u$, we observe a timestamped interaction sequence
\begin{equation}
\mathcal{S}_u = [(v_1,t_1),(v_2,t_2),\ldots,(v_L,t_L)],
\end{equation}
where interactions are sorted in chronological order and $v_l \in \mathcal{V}$ denotes the item interacted with at time $t_l$. 
The goal of sequential recommendation is to predict the next item $v_{L+1}$ given the historical timestamped sequence $\mathcal{S}_u$.

In SID-based generative recommendation, each item is represented by a compact sequence of discrete semantic codes. 
Formally, a semantic ID encoder maps an item $v$ to a length-$K$ code tuple
\begin{equation}
\Phi(v) = (c_v^1,c_v^2,\ldots,c_v^K),
\end{equation}
where $c_v^k \in \mathcal{C}_k$ is the code at level $k$ and $\mathcal{C}_k=\{1,\ldots,B_k\}$ denotes the corresponding codebook. 
Instead of predicting $v_{L+1}$ directly from the full item vocabulary, the generative recommender predicts the target item's semantic ID $\Phi(v_{L+1})$ autoregressively and then maps the generated code tuple back to items through a lookup table.

To incorporate temporal information, we define inter-interaction gaps between consecutive historical interactions. 
For $l \ge 2$, the elapsed time before the $l$-th interaction is
\begin{equation}
\Delta t_l = t_l - t_{l-1}.
\end{equation}
We discretize $\Delta t_l$ into a symbolic gap token
\begin{equation}
g_l = \textsc{Disc}(\Delta t_l),
\end{equation}
where $\textsc{Disc}(\cdot)$ maps continuous time intervals into $G$ predefined temporal bins. 
In this work, we use $G=5$ real gap bins with log-scale thresholds
$\{1\mathrm{h},1\mathrm{d},1\mathrm{w},1\mathrm{mo}\}$, corresponding to
$<1$h, $[1\mathrm{h},1\mathrm{d})$, $[1\mathrm{d},1\mathrm{w})$, $[1\mathrm{w},1\mathrm{mo})$, and $\ge 1$mo. 
Since the first interaction has no previous timestamp, we assign it a special start-of-history token $g_{\mathrm{start}}$ and set
\begin{equation}
g_1 = g_{\mathrm{start}}.
\end{equation}

Given these definitions, ChronoSID constructs a temporally augmented encoder input by interleaving gap tokens with item semantic IDs:
\begin{equation}
\mathbf{x}_u =
\bigl[
g_1,\, \Phi(v_1),\,
g_2,\, \Phi(v_2),\,
\ldots,\,
g_L,\, \Phi(v_L)
\bigr].
\end{equation}
The learning objective is to model the conditional probability
\begin{equation}
p_\theta\!\left(\Phi(v_{L+1}) \mid \mathbf{x}_u\right),
\end{equation}
where the target semantic ID is generated token by token. 
At inference time, candidate semantic ID tuples are produced by autoregressive decoding and then resolved to items using the semantic ID lookup table to obtain the final ranked recommendation list.

Compared with ReSID, which represents the input history as
\begin{equation}
\mathbf{x}^{\mathrm{ReSID}}_u =
[\Phi(v_1),\Phi(v_2),\ldots,\Phi(v_L)],
\end{equation}
ChronoSID explicitly augments the historical SID sequence with inter-interaction gap tokens. 
This formulation preserves the compact discrete-code generation paradigm while allowing the sequence generator to exploit temporal cues about user behavior continuity and drift.

\section{Methodology}
\label{sec:method}

\begin{figure*}[t]
  \centering
  \includegraphics[width=\linewidth]{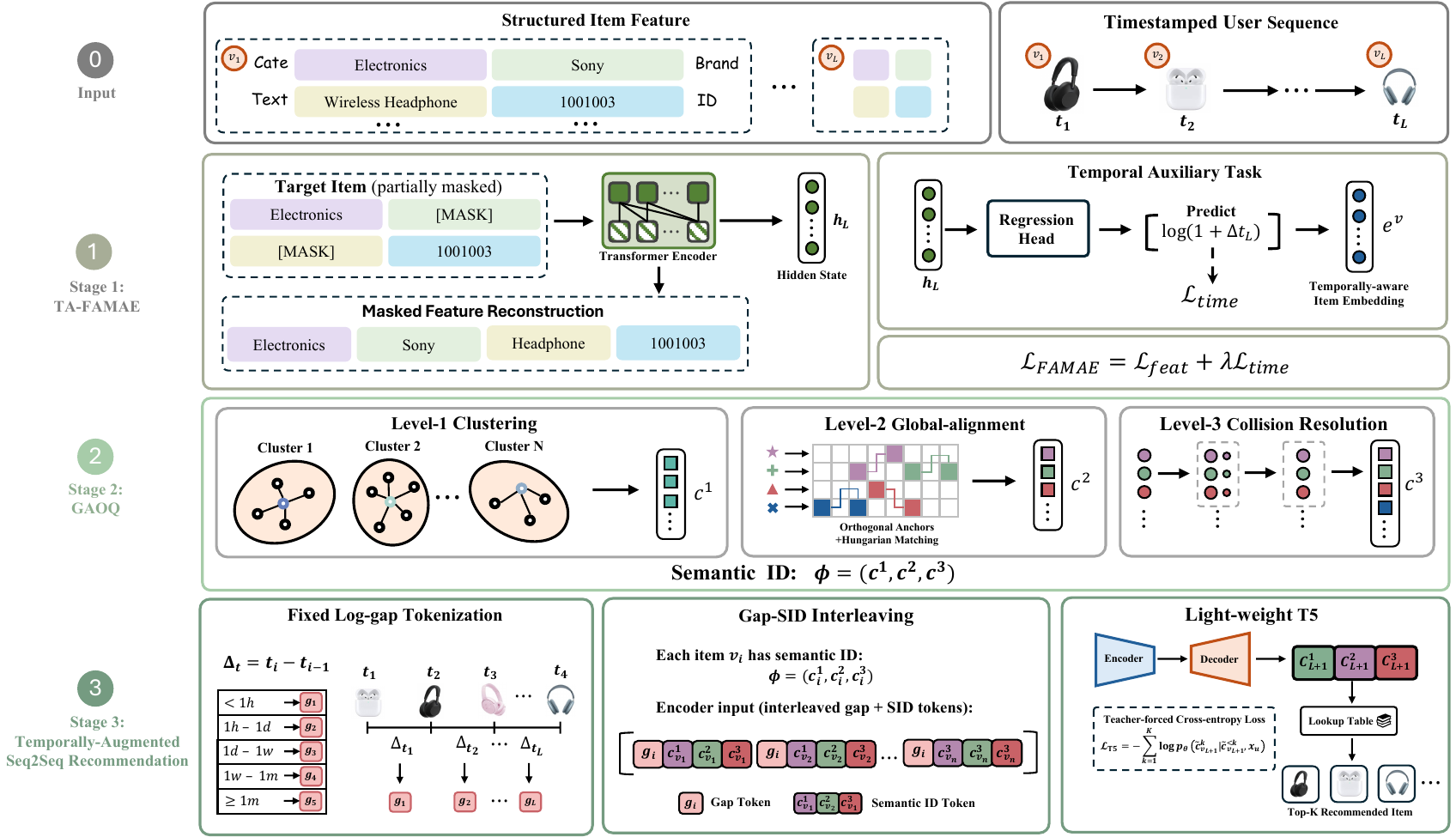}
  \vspace{-2em}
  \caption{
  Overview of ChronoSID. 
  Stage~1 learns item representations with Time-Aware Field-Aware Masked Auto-Encoding (TA-FAMAE), which combines masked feature reconstruction with an auxiliary time-gap prediction objective. 
  Stage~2 adopts Globally Aligned Orthogonal Quantization (GAOQ) to convert the frozen item representations into three-level semantic ID tuples. 
  Stage~3 discretizes historical inter-interaction intervals into fixed log-scale gap tokens and interleaves them with semantic ID tokens as the T5 encoder input for temporally augmented sequence-to-sequence recommendation.
  }
  \label{fig:overview}
\end{figure*}

ChronoSID follows a three-stage pipeline for SID-based generative sequential recommendation, as illustrated in Figure~\ref{fig:overview}. 
Given structured item features and timestamped user interaction sequences, ChronoSID first learns item representations with temporal regularization, then quantizes the learned representations into compact semantic ID tuples, and finally trains a sequence-to-sequence generator over temporally augmented SID sequences. 
The overall design builds on the ReSID pipeline~\cite{resid}: Stage~2 keeps the same quantization procedure for a controlled comparison, while Stages~1 and~3 introduce temporal information at the representation-learning and sequence-modeling levels, respectively.

\subsection{Stage 1: Time-Aware Item Representation (TA-FAMAE)}
\label{sec:famae}

ChronoSID first learns item representations from structured recommendation features.
We build upon the field-aware masked auto-encoding objective used in ReSID and introduce an auxiliary temporal prediction task.
The purpose of this stage is not to make item identifiers dynamic, but to regularize item representations with temporal predictability before semantic quantization.

\subsubsection{Feature-Aware Masked Encoding.}

Each item $v$ is associated with $F$ structured feature fields,
\begin{equation}
\mathbf{f}_v = (f_v^1,f_v^2,\ldots,f_v^F),
\end{equation}
where each field may correspond to categorical, textual, or identifier-based item information.
For a training instance ending at target item $v_L$, we use its preceding interactions as context and randomly mask a subset of the target item's feature fields.
Specifically, we sample a non-empty mask set $\mathcal{M}\subseteq \{1,\ldots,F\}$ and replace $\{f_{v_L}^f\}_{f\in\mathcal{M}}$ with learnable \texttt{[MASK]} tokens.

A Transformer encoder $\mathcal{E}$ processes the historical context together with the partially masked target item and produces a contextual representation $\mathbf{h}_L\in\mathbb{R}^d$ at the target position.
For each feature field $f$, we maintain a field-specific embedding table
$\mathbf{E}_f\in\mathbb{R}^{|\mathcal{V}_f|\times d}$, where $\mathcal{V}_f$ is the vocabulary of field $f$.
The masked feature value is reconstructed by comparing $\mathbf{h}_L$ with candidate embeddings in the corresponding field space:
\begin{equation}
\hat{p}_f(\cdot \mid \mathbf{h}_L)
=
\mathrm{softmax}
\left(
s_f \cdot
\frac{\mathbf{h}_L \mathbf{E}_f^\top}
{\|\mathbf{h}_L\|\,\|\mathbf{E}_f\|}
\right),
\label{eq:cosine_sim}
\end{equation}
where $s_f$ is a learnable temperature parameter for feature field $f$.
The masked feature reconstruction loss is then defined as
\begin{equation}
\mathcal{L}_{\mathrm{feat}}
=
\sum_{f\in\mathcal{M}}
w_f\,
\mathcal{L}_{\mathrm{CE}}
\left(
\hat{p}_f(\cdot \mid \mathbf{h}_L),
f_{v_L}^f
\right),
\label{eq:feat_loss}
\end{equation}
where $w_f$ is a field-specific loss weight.
At each training step, the number of masked fields is sampled uniformly from $\{1,\ldots,F\}$, encouraging the encoder to learn robust item representations under partial feature observations.

\subsubsection{Temporal Auxiliary Task.}

To incorporate temporal regularization into item representation learning, we add an auxiliary task that predicts the elapsed time between the target item and its immediately preceding interaction.
For the target position $L$, the time interval is
\begin{equation}
\Delta t_L = t_L - t_{L-1}.
\end{equation}
We apply a linear regression head to the target representation and predict the log-transformed interval:
\begin{equation}
\hat{z}_L = \mathbf{w}_g^\top \mathbf{h}_L + b_g,
\end{equation}
\begin{equation}
\mathcal{L}_{\mathrm{time}}
=
\left(
\hat{z}_L - \log(1+\Delta t_L)
\right)^2.
\label{eq:time_loss}
\end{equation}
The overall TA-FAMAE objective is
\begin{equation}
\mathcal{L}_{\mathrm{TA\text{-}FAMAE}}
=
\mathcal{L}_{\mathrm{feat}}
+
\lambda \mathcal{L}_{\mathrm{time}},
\label{eq:famae_loss}
\end{equation}
where $\lambda$ controls the strength of temporal regularization and is set to $0.1$ by default.
This auxiliary objective encourages the target-position representation to preserve temporal predictability in addition to feature semantics.
After training, we extract one deterministic embedding $\mathbf{e}_v$ for each item $v$ using the trained feature-aware encoder with the item's unmasked feature fields, and freeze these embeddings for semantic quantization.

\subsection{Stage 2: Globally Aligned Orthogonal Quantization (GAOQ)}
\label{sec:gaoq}

Given the frozen item embeddings $\{\mathbf{e}_v\}_{v\in\mathcal{V}}$, ChronoSID adopts the Globally Aligned Orthogonal Quantization (GAOQ) procedure from ReSID~\cite{resid} to obtain compact semantic ID tuples.
We keep this quantization stage unchanged so that the comparison with ReSID isolates the effect of temporal modeling introduced in Stages~1 and~3.
GAOQ assigns each item a three-level semantic ID,
\begin{equation}
\Phi(v) = (c_v^1,c_v^2,c_v^3),
\end{equation}
where the first level captures coarse item groups, the second level aligns sub-cluster semantics across coarse groups, and the third level resolves collisions to ensure unique item identifiers.

\subsubsection{Level-1 Clustering.}

At the first level, balanced K-Means~\cite{malinen2014balanced} partitions item embeddings into $b_1$ coarse clusters with approximately balanced sizes.
Each item is assigned a first-level code
\begin{equation}
c_v^1 \in \{1,\ldots,b_1\}.
\end{equation}
Following ReSID, we use $b_1=32$ in our experiments.

\subsubsection{Level-2 Clustering with Global Alignment.}

Within each Level-1 cluster, GAOQ further applies K-Means to obtain $b_2$ sub-clusters.
A direct use of local sub-cluster indices would introduce permutation ambiguity, because the same sub-cluster index may correspond to different residual directions in different Level-1 clusters.
To reduce this ambiguity, GAOQ aligns local sub-clusters to a set of globally shared orthogonal anchors.

Let $\boldsymbol{\mu}^{(i)}$ be the centroid of Level-1 cluster $i$, and let
$\{\boldsymbol{\mu}_j^{(i)}\}_{j=1}^{b_2}$ be the centroids of its Level-2 sub-clusters.
We construct $g_2=b_2$ orthogonal anchor vectors
\begin{equation}
\mathbf{Q}_2 = [\mathbf{q}_1;\mathbf{q}_2;\ldots;\mathbf{q}_{g_2}]
\in\mathbb{R}^{g_2\times d},
\end{equation}
where the rows are obtained from a random orthonormal matrix.
For each Level-1 cluster $i$, the residual direction of sub-cluster $j$ is
\begin{equation}
\mathbf{r}_j^{(i)}
=
\boldsymbol{\mu}_j^{(i)}-\boldsymbol{\mu}^{(i)}.
\end{equation}
The local sub-clusters are matched to the global anchors by solving
\begin{equation}
\sigma_i^*
=
\operatorname*{arg\,min}_{\sigma\in\mathfrak{S}_{b_2}}
\sum_{j=1}^{b_2}
\left(
1-
\cos\left(
\mathbf{r}_j^{(i)},\mathbf{q}_{\sigma(j)}
\right)
\right),
\label{eq:hungarian}
\end{equation}
where $\mathfrak{S}_{b_2}$ denotes the set of all permutations over $b_2$ elements.
The aligned second-level code of item $v$ in Level-1 cluster $i$ is then
\begin{equation}
c_v^2
=
\sigma_i^*
\left(
\mathrm{subcluster}(v)
\right).
\end{equation}
This alignment encourages the same second-level code to represent similar residual directions across different coarse clusters.

\subsubsection{Level-3 Collision Resolution.}

Items that share the same $(c_v^1,c_v^2)$ prefix are further disambiguated at the third level.
Let $n_{\max}$ be the maximum number of items under any $(c^1,c^2)$ prefix.
GAOQ sets the third-level code capacity to $b_3=n_{\max}$ and assigns third-level codes within each prefix group.
A second set of orthogonal anchors is constructed analogously, and a Hungarian matching step is applied to assign items to third-level codes.
The resulting three-level semantic ID is unique for each item and is used as the target sequence for generative recommendation.

\subsection{Stage 3: Temporally-Augmented Seq2Seq Recommendation}
\label{sec:t5}

After obtaining semantic IDs, ChronoSID trains a sequence-to-sequence generator for next-item prediction.
The key difference from ReSID is that ChronoSID augments the encoder input with temporal gap tokens, while keeping the decoder target as the next item's semantic ID.

\subsubsection{Vocabulary Construction.}

The three code levels from GAOQ are first remapped to globally unique token indices to avoid collisions across code levels.
Let $\rho_k(\cdot)$ denote the remapping function for code level $k$, and define
\begin{equation}
\tilde{c}_v^k = \rho_k(c_v^k).
\end{equation}
All remapped semantic ID tokens are contained in $\{1,\ldots,M\}$, where $M$ is the maximum remapped code index.

We reserve token $0$ for \texttt{[PAD]} and token $M+1$ for \texttt{[EOS]}.
For temporal information, ChronoSID uses $G=5$ real gap bins and one special start-of-history token.
Let
\begin{equation}
\mathcal{G}_{\mathrm{tok}}
=
\{g_{\mathrm{start}}, g^{(1)}, g^{(2)}, \ldots, g^{(G)}\}.
\end{equation}
Here, $\mathcal{G}_{\mathrm{tok}}$ denotes the temporal token set.
The complete T5 vocabulary is
\begin{equation}
\mathcal{W}
=
\underbrace{\{0\}}_{\texttt{[PAD]}}
\cup
\underbrace{\{1,\ldots,M\}}_{\text{semantic ID tokens}}
\cup
\underbrace{\{M+1\}}_{\texttt{[EOS]}}
\cup
\underbrace{\{M+2,\ldots,M+G+2\}}_{\text{gap tokens}}.
\label{eq:vocab}
\end{equation}
We define a mapping $\psi(\cdot)$ from temporal tokens to vocabulary indices:
\begin{equation}
\psi(g_{\mathrm{start}})=M+2,
\end{equation}
\begin{equation}
\psi(g^{(r)})=M+2+r,\quad r=1,\ldots,G.
\end{equation}

\subsubsection{Gap Token Injection.}

For each historical interaction position $l$, we assign a temporal token before the item's semantic ID tuple.
For the first interaction, we use the start-of-history token:
\begin{equation}
g_1 = g_{\mathrm{start}}.
\end{equation}
For $l\ge 2$, the real-valued time interval is discretized into one of the $G$ log-scale bins:
\begin{equation}
g_l = g^{(\textsc{Disc}(t_l-t_{l-1}))}.
\end{equation}
The corresponding vocabulary-level temporal token is
\begin{equation}
\delta_l = \psi(g_l).
\end{equation}

The temporally augmented encoder input is constructed by interleaving each gap token with the remapped semantic ID tuple of the corresponding item:
\begin{equation}
\mathbf{x}_u =
\bigl[
\delta_1,\tilde{c}_{v_1}^1,\tilde{c}_{v_1}^2,\tilde{c}_{v_1}^3,\;
\delta_2,\tilde{c}_{v_2}^1,\tilde{c}_{v_2}^2,\tilde{c}_{v_2}^3,\;
\ldots,\;
\delta_L,\tilde{c}_{v_L}^1,\tilde{c}_{v_L}^2,\tilde{c}_{v_L}^3
\bigr].
\label{eq:input_seq}
\end{equation}
Thus, each historical item contributes one temporal token and three semantic ID tokens, resulting in an encoder length of $4L$.
By contrast, ReSID uses only the semantic ID tuples and has encoder length $3L$.
The gap tokens are used only as encoder-side temporal context; the decoder still generates semantic ID tokens of the target item.

\subsubsection{Training and Inference.}

ChronoSID uses a lightweight T5-style encoder-decoder model \cite{raffel2020exploring} trained from scratch.
Given the temporally augmented encoder input $\mathbf{x}_u$, the decoder predicts the target item's semantic ID:
\begin{equation}
\Phi(v_{L+1})
=
(\tilde{c}_{v_{L+1}}^1,\tilde{c}_{v_{L+1}}^2,\tilde{c}_{v_{L+1}}^3).
\end{equation}
The training objective is teacher-forced cross-entropy:
\begin{equation}
\mathcal{L}_{\mathrm{T5}}
=
-\sum_{k=1}^{K}
\log
p_\theta
\left(
\tilde{c}_{v_{L+1}}^k
\mid
\tilde{c}_{v_{L+1}}^{<k},\mathbf{x}_u
\right),
\label{eq:t5_loss}
\end{equation}
where $K=3$ in our implementation.
Since the decoder target contains only semantic ID tokens, gap tokens serve purely as temporal conditioning signals for the encoder.

During inference, the model autoregressively generates semantic ID tuples with beam search.
We use per-step beam sizes $[B_1,B_2,B_3]=[50,50,50]$.
Generated tuples are resolved through the semantic ID lookup table; unmatched tuples are discarded, and duplicated items are removed.
The remaining items are ranked according to their generation scores to produce the final Top-$K$ recommendation list.

\section{Experiments}
\label{sec:experiments}

We conduct experiments to evaluate the effectiveness, robustness, and efficiency of ChronoSID.
Specifically, we organize our experiments around the following research questions:

\begin{itemize}[leftmargin=*] 
  \item[\textbf{RQ1}]
  How does ChronoSID perform compared with representative sequential recommenders and recent SID-based generative recommendation methods across different product domains?

  \item[\textbf{RQ2}]
  How much does each temporal component contribute to the final performance?
  In particular, what are the effects of the temporal auxiliary objective in TA-FAMAE and gap-token interleaving in the sequence-to-sequence generator?

  \item[\textbf{RQ3}]
  Does ChronoSID improve recommendation performance under long-gap scenarios, where short-term interest continuity becomes weaker and preference drift is more likely to occur?

  \item[\textbf{RQ4}]
  How sensitive is ChronoSID to key temporal hyperparameters, including the temporal loss weight $\lambda$ and the number of discretized gap bins $G$?

  \item[\textbf{RQ5}]
  What computational overhead does ChronoSID introduce compared with ReSID, considering that it increases the encoder input length while keeping the decoder target unchanged?

  \item[\textbf{RQ6}]
  Does ChronoSID improve recommendations for both popular and less popular target items, or are its gains mainly driven by popularity effects?

  \item[\textbf{RQ7}]
  Does temporal augmentation help the generator produce more accurate semantic ID candidates under the same decoding and lookup protocol?
\end{itemize}

\begin{table}[t]
\centering
\caption{Dataset statistics.}
\vspace{-1em}
\label{tab:dataset_stats}
\begin{tabular}{lrrrr}
\toprule
\textbf{Data} & \textbf{\#Users} & \textbf{\#Items} & \textbf{\#Inter.} & \textbf{Density} \\
\midrule
MI  & 57,359  & 23,742  & 490,522   & 0.036\% \\
VG  & 94,515  & 24,685  & 772,218   & 0.033\% \\
IS  & 50,886  & 25,142  & 394,989   & 0.031\% \\
BP  & 150,642 & 35,024  & 1,189,171 & 0.023\% \\
ACS & 196,980 & 87,449  & 1,706,484 & 0.010\% \\
SO  & 409,309 & 151,411 & 3,333,753 & 0.005\% \\
TG  & 431,411 & 156,537 & 3,652,250 & 0.005\% \\
BPC & 712,259 & 193,383 & 5,785,124 & 0.004\% \\
\bottomrule
\end{tabular}
\vspace{-0.8em}
\end{table}

\begin{table*}[htbp]
\centering
\caption{
Full main results. Datasets: MI = Musical Instruments, VG = Video Games, 
IS = Industrial \& Scientific, BP = Baby Products, ACS = Arts, Crafts \& Sewing, 
SO = Sports \& Outdoors, TG = Toys \& Games, and BPC = Beauty \& Personal Care. 
For each dataset, we report Recall (R@5, R@10) and NDCG (N@5, N@10). 
The best result in each column is shown in bold, and the second-best result is underlined. 
Models marked with $^{\ast}$, such as HGN$^{\ast}$, are augmented with side-information fields.
}
\label{tab:main_results}
\scriptsize
\setlength{\tabcolsep}{4.0pt}
\renewcommand{\arraystretch}{1.05}
\resizebox{\textwidth}{!}{
\begin{tabular}{lcccccccccccccccc}
\toprule
\multicolumn{17}{c}{\textbf{(A) MI / VG / IS / BP}} \\
\midrule
\multirow{2}{*}{Model}
& \multicolumn{4}{c}{MI}
& \multicolumn{4}{c}{VG}
& \multicolumn{4}{c}{IS}
& \multicolumn{4}{c}{BP} \\
\cmidrule(lr){2-5} \cmidrule(lr){6-9} \cmidrule(lr){10-13} \cmidrule(lr){14-17}
& R@5 & R@10 & N@5 & N@10
& R@5 & R@10 & N@5 & N@10
& R@5 & R@10 & N@5 & N@10
& R@5 & R@10 & N@5 & N@10 \\
\midrule
HGN              & 0.0309 & 0.0498 & 0.0192 & 0.0253 & 0.0408 & 0.0685 & 0.0248 & 0.0337 & 0.0214 & 0.0361 & 0.0132 & 0.0179 & 0.0164 & 0.0276 & 0.0101 & 0.0137 \\
SASRec           & 0.0320 & 0.0528 & 0.0199 & 0.0265 & 0.0509 & 0.0820 & 0.0322 & 0.0422 & 0.0216 & 0.0349 & 0.0133 & 0.0176 & 0.0206 & 0.0338 & 0.0131 & 0.0174 \\
BERT4Rec          & 0.0324 & 0.0517 & 0.0206 & 0.0268 & 0.0479 & 0.0783 & 0.0303 & 0.0401 & 0.0207 & 0.0341 & 0.0131 & 0.0174 & 0.0204 & 0.0338 & 0.0129 & 0.0172 \\
S$^3$-Rec        & 0.0336 & 0.0543 & 0.0212 & 0.0278 & 0.0495 & 0.0801 & 0.0315 & 0.0413 & 0.0210 & 0.0349 & 0.0135 & 0.0179 & 0.0213 & 0.0360 & 0.0135 & 0.0182 \\
\midrule
HGN$^{*}$        & 0.0287 & 0.0469 & 0.0183 & 0.0242 & 0.0431 & 0.0541 & 0.0212 & 0.0280 & 0.0192 & 0.0321 & 0.0122 & 0.0163 & 0.0169 & 0.0278 & 0.0108 & 0.0143 \\
SASRec$^{*}$     & 0.0397 & \underline{0.0639} & 0.0248 & 0.0326 & 0.0532 & \underline{0.0914} & 0.0295 & 0.0417 & \underline{0.0333} & \textbf{0.0538} & 0.0185 & 0.0250 & 0.0257 & \underline{0.0425} & 0.0157 & 0.0211 \\
BERT4Rec$^{*}$    & 0.0377 & 0.0614 & 0.0241 & 0.0317 & 0.0537 & 0.0879 & 0.0337 & 0.0447 & 0.0296 & 0.0488 & 0.0186 & 0.0248 & 0.0241 & 0.0390 & 0.0155 & 0.0202 \\
S$^3$-Rec$^{*}$  & 0.0380 & 0.0611 & 0.0242 & 0.0317 & 0.0516 & 0.0843 & 0.0326 & 0.0431 & 0.0274 & 0.0441 & 0.0169 & 0.0223 & 0.0237 & 0.0386 & 0.0152 & 0.0200 \\
\midrule
TIGER            & 0.0385 & 0.0592 & 0.0251 & 0.0318 & 0.0456 & 0.0882 & 0.0374 & 0.0475 & 0.0297 & 0.0460 & 0.0192 & 0.0244 & 0.0256 & 0.0406 & 0.0160 & 0.0217 \\
LETTER           & \underline{0.0406} & 0.0623 & \underline{0.0269} & \underline{0.0338} & \underline{0.0592} & \underline{0.0914} & \underline{0.0389} & \underline{0.0493} & 0.0307 & 0.0482 & \underline{0.0197} & \underline{0.0254} & 0.0261 & 0.0416 & 0.0171 & 0.0221 \\
EAGER            & 0.0327 & 0.0523 & 0.0210 & 0.0273 & 0.0547 & 0.0863 & 0.0357 & 0.0458 & 0.0252 & 0.0406 & 0.0163 & 0.0212 & 0.0199 & 0.0331 & 0.0126 & 0.0168 \\
UNGER             & 0.0362 & 0.0567 & 0.0233 & 0.0299 & 0.0547 & 0.0859 & 0.0354 & 0.0454 & 0.0232 & 0.0374 & 0.0150 & 0.0196 & 0.0221 & 0.0355 & 0.0144 & 0.0187 \\
ETEGRec            & 0.0372 & 0.0579 & 0.0244 & 0.0310 & 0.0558 & 0.0869 & 0.0367 & 0.0466 & 0.0251 & 0.0393 & 0.0166 & 0.0212 & 0.0229 & 0.0369 & 0.0151 & 0.0196 \\
\midrule
ReSID            & 0.0388 & 0.0614 & 0.0253 & 0.0325 & 0.0571 & 0.0898 & 0.0375 & 0.0480 & 0.0305 & 0.0478 & 0.0194 & 0.0250 & \underline{0.0272} & 0.0422 & \underline{0.0179} & \underline{0.0227} \\
ChronoSID        & \textbf{0.0417} & \textbf{0.0645} & \textbf{0.0273} & \textbf{0.0346} & \textbf{0.0597} & \textbf{0.0927} & \textbf{0.0396} & \textbf{0.0501} & \textbf{0.0340} & \underline{0.0512} & \textbf{0.0218} & \textbf{0.0273} & \textbf{0.0285} & \textbf{0.0441} & \textbf{0.0186} & \textbf{0.0236} \\
\midrule
\multicolumn{17}{c}{\textbf{(B) ACS / SO / TG / BPC}} \\
\midrule
\multirow{2}{*}{Model}
& \multicolumn{4}{c}{ACS}
& \multicolumn{4}{c}{SO}
& \multicolumn{4}{c}{TG}
& \multicolumn{4}{c}{BPC} \\
\cmidrule(lr){2-5} \cmidrule(lr){6-9} \cmidrule(lr){10-13} \cmidrule(lr){14-17}
& R@5 & R@10 & N@5 & N@10
& R@5 & R@10 & N@5 & N@10
& R@5 & R@10 & N@5 & N@10
& R@5 & R@10 & N@5 & N@10 \\
\midrule
HGN              & 0.0190 & 0.0314 & 0.0119 & 0.0159 & 0.0112 & 0.0184 & 0.0070 & 0.0093 & 0.0129 & 0.0214 & 0.0077 & 0.0105 & 0.0123 & 0.0207 & 0.0076 & 0.0103 \\
SASRec           & 0.0218 & 0.0353 & 0.0138 & 0.0182 & 0.0140 & 0.0231 & 0.0089 & 0.0118 & 0.0169 & 0.0270 & 0.0105 & 0.0138 & 0.0157 & 0.0254 & 0.0101 & 0.0132 \\
BERT4Rec          & 0.0211 & 0.0343 & 0.0134 & 0.0176 & 0.0131 & 0.0216 & 0.0082 & 0.0110 & 0.0159 & 0.0255 & 0.0101 & 0.0131 & 0.0150 & 0.0246 & 0.0095 & 0.0126 \\
S$^3$-Rec        & 0.0228 & 0.0370 & 0.0145 & 0.0191 & 0.0153 & 0.0249 & 0.0097 & 0.0128 & 0.0186 & 0.0297 & 0.0117 & 0.0153 & 0.0171 & 0.0282 & 0.0110 & 0.0145 \\
\midrule
HGN$^{*}$        & 0.0166 & 0.0271 & 0.0107 & 0.0141 & 0.0121 & 0.0197 & 0.0078 & 0.0102 & 0.0119 & 0.0193 & 0.0075 & 0.0098 & 0.0137 & 0.0219 & 0.0088 & 0.0115 \\
SASRec$^{*}$     & 0.0275 & 0.0464 & 0.0157 & 0.0218 & \underline{0.0197} & \textbf{0.0319} & 0.0122 & 0.0161 & \textbf{0.0243} & \textbf{0.0401} & 0.0127 & 0.0178 & \underline{0.0208} & \underline{0.0340} & 0.0126 & \underline{0.0169} \\
BERT4Rec$^{*}$    & 0.0275 & 0.0439 & 0.0176 & 0.0229 & 0.0175 & 0.0286 & 0.0111 & 0.0147 & 0.0223 & \underline{0.0356} & 0.0138 & 0.0181 & 0.0192 & 0.0312 & 0.0123 & 0.0161 \\
S$^3$-Rec$^{*}$  & 0.0261 & 0.0420 & 0.0168 & 0.0219 & 0.0184 & 0.0297 & 0.0117 & 0.0154 & 0.0209 & 0.0338 & 0.0128 & 0.0170 & 0.0198 & 0.0322 & 0.0125 & 0.0165 \\
\midrule
TIGER            & 0.0285 & 0.0446 & 0.0187 & 0.0239 & 0.0180 & 0.0279 & 0.0119 & 0.0151 & 0.0214 & 0.0322 & 0.0141 & 0.0176 & 0.0192 & 0.0299 & 0.0125 & 0.0160 \\
LETTER           & 0.0297 & 0.0464 & 0.0196 & 0.0249 & 0.0187 & 0.0288 & 0.0124 & 0.0156 & 0.0221 & 0.0340 & 0.0147 & 0.0185 & 0.0203 & 0.0315 & \underline{0.0133} & \underline{0.0169} \\
EAGER            & 0.0240 & 0.0375 & 0.0155 & 0.0198 & 0.0145 & 0.0236 & 0.0095 & 0.0125 & 0.0211 & 0.0289 & 0.0127 & 0.0159 & 0.0164 & 0.0265 & 0.0107 & 0.0139 \\
UNGER             & 0.0253 & 0.0395 & 0.0164 & 0.0210 & 0.0164 & 0.0260 & 0.0109 & 0.0140 & 0.0200 & 0.0302 & 0.0132 & 0.0165 & 0.0182 & 0.0286 & 0.0119 & 0.0152 \\
ETEGRec            & 0.0249 & 0.0389 & 0.0162 & 0.0206 & 0.0148 & 0.0236 & 0.0097 & 0.0125 & 0.0178 & 0.0270 & 0.0116 & 0.0146 & 0.0177 & 0.0274 & 0.0115 & 0.0146 \\
\midrule
ReSID            & \underline{0.0300} & \underline{0.0468} & \underline{0.0198} & \underline{0.0252} & 0.0195 & 0.0305 & \underline{0.0128} & \underline{0.0164} & 0.0233 & 0.0352 & \underline{0.0153} & \underline{0.0191} & 0.0223 & 0.0346 & 0.0146 & 0.0185 \\
ChronoSID        & \textbf{0.0309} & \textbf{0.0475} & \textbf{0.0202} & \textbf{0.0255} & \textbf{0.0199} & \underline{0.0309} & \textbf{0.0130} & \textbf{0.0166} & \underline{0.0235} & 0.0355 & \textbf{0.0154} & \textbf{0.0193} & \textbf{0.0228} & \textbf{0.0348} & \textbf{0.0149} & \textbf{0.0189} \\
\bottomrule
\end{tabular}
}
\end{table*}

\textbf{Datasets.}
We evaluate ChronoSID on eight subsets of the Amazon-2023 review dataset~\cite{amazon2023}: 
\textit{Musical Instruments} (MI), \textit{Video Games} (VG), \textit{Industrial \& Scientific} (IS), \textit{Baby Products} (BP), \textit{Arts, Crafts \& Sewing} (ACS), \textit{Sports \& Outdoors} (SO), \textit{Toys \& Games} (TG), and \textit{Beauty \& Personal Care} (BPC). The statistics of the processed datasets are summarized in Table~\ref{tab:dataset_stats}.
Following prior work on generative recommendation~\cite{letter,eager,etegrec,resid}, we sort each user's interactions chronologically and discard sparse users and items with fewer than five interactions.
We then adopt the leave-one-out evaluation protocol: the last interaction of each user is used for testing, the second-to-last interaction is used for validation, and the remaining interactions are used for training.
Each item is associated with structured fields such as title, category, brand, and item identifier, which are used for item representation learning.
Timestamps are used only to compute historical inter-interaction gaps for temporal tokens and the auxiliary temporal objective.

\textbf{Baselines.}
We compare ChronoSID with representative sequential recommendation models and recent SID-based generative recommendation methods.

\noindent\textbf{Sequential recommendation methods:}
\begin{itemize}
    \item \textbf{HGN}~\cite{hgn}: a hierarchical gating network that captures both short-term and long-term user preferences from interaction sequences.
    \item \textbf{SASRec}~\cite{sasrec}: a unidirectional Transformer-based sequential recommender that models item dependencies with self-attention.
    \item \textbf{BERT4Rec}~\cite{bert4rec}: a bidirectional Transformer model that learns contextualized sequence representations through masked item prediction.
    \item \textbf{S$^3$-Rec}~\cite{s3rec}: a self-supervised sequential recommender that pretrains item and sequence representations with multiple mutual-information-based objectives before fine-tuning for next-item prediction.
\end{itemize}

\noindent\textbf{Generative recommendation methods:}
\begin{itemize}
    \item \textbf{TIGER}~\cite{tiger}: a pioneering SID-based generative recommender that obtains item identifiers with RQ-VAE quantization and generates target item IDs autoregressively.
    \item \textbf{LETTER}~\cite{letter}: a learnable item tokenization method that improves semantic identifiers by incorporating hierarchical semantics, collaborative signals, and code assignment diversity.
    \item \textbf{EAGER}~\cite{eager}: a two-stream generative recommender that separately models semantic and behavioral information for behavior-semantic collaboration.
    \item \textbf{UNGER}~\cite{unger}: a unified-code generative recommendation method that integrates semantic and collaborative signals into item identifier learning.
    \item \textbf{ETEGRec}~\cite{etegrec}: an end-to-end framework that jointly optimizes item tokenization and generative recommendation.
    \item \textbf{ReSID}~\cite{resid}: a recommendation-native semantic ID framework that learns item representations from structured recommendation features and applies globally aligned orthogonal quantization to produce compact and predictable item code sequences. ReSID is the most direct baseline for ChronoSID, since ChronoSID follows the same three-stage pipeline and keeps the GAOQ quantization stage unchanged.
\end{itemize}
The starred variants denote side-information-enhanced versions of conventional sequential recommenders, where the same structured item fields are incorporated following the ReSID setting.
For a controlled comparison with ReSID, ChronoSID uses the same structured item features, semantic ID depth, quantization configuration, generator backbone, and beam search setting unless otherwise specified.

\textbf{Evaluation settings.}
Following prior work~\cite{tiger,letter,etegrec}, we evaluate recommendation performance using Recall@K and NDCG@K with $K \in \{5,10\}$.
We adopt the leave-one-out protocol: for each user, the last interaction is used for testing, the second-to-last interaction is used for validation, and all earlier interactions are used for training.
For conventional sequential recommenders, items are ranked according to their predicted relevance scores.
For SID-based generative recommenders, the decoder generates candidate semantic ID tuples via beam search, which are then mapped back to items through the semantic ID lookup table.
Generated tuples that do not correspond to any item are discarded, and duplicated items are merged by keeping the highest generation score.
The final ranked list is evaluated against the held-out test item.
All hyperparameters are selected based on validation performance, and test results are reported only after model selection.

\textbf{Implementation details.} For ChronoSID, we follow ReSID~\cite{resid} to use three-level semantic IDs, i.e., $K=3$.
Unless otherwise specified, we set the number of real temporal gap bins to $G=5$ with log-scale thresholds $\{1\mathrm{h},1\mathrm{d},1\mathrm{w},1\mathrm{mo}\}$, and set the temporal auxiliary loss weight to $\lambda=0.1$.
The sequence generator is a lightweight T5-style encoder--decoder trained from scratch.
During inference, SID candidates are generated with per-level beam sizes $[50,50,50]$.
For efficiency evaluation, all runtime measurements are conducted on the same hardware with an identical batch size, beam size, and evaluation protocol. The main results in Table~\ref{tab:main_results} are averaged over five runs with different random seeds. All experiments are conducted on NVIDIA V100 GPUs under the same hardware settings.

\begin{table*}[t]
\centering
\caption{Ablation study results on MI, VG, and IS datasets. Bold values denote the best results.}
\label{tab:ablation}
\small
\setlength{\tabcolsep}{6pt}
\begin{tabular}{l|cccc|cccc|cccc}
\toprule
\multirow{2}{*}{\textbf{Methods}} 
& \multicolumn{4}{c|}{\textbf{MI}} 
& \multicolumn{4}{c|}{\textbf{VG}} 
& \multicolumn{4}{c}{\textbf{IS}} \\
\cmidrule(lr){2-5} \cmidrule(lr){6-9} \cmidrule(lr){10-13}
& \textbf{R@5} & \textbf{R@10} & \textbf{N@5} & \textbf{N@10} 
& \textbf{R@5} & \textbf{R@10} & \textbf{N@5} & \textbf{N@10}
& \textbf{R@5} & \textbf{R@10} & \textbf{N@5} & \textbf{N@10} \\
\midrule
ReSID             
& 0.0388 & 0.0614 & 0.0253 & 0.0325 
& 0.0571 & 0.0898 & 0.0375 & 0.0480
& 0.0305 & 0.0478 & 0.0194 & 0.0250 \\

+ TA-FAMAE
& 0.0401 & 0.0620 & 0.0262 & 0.0333 
& 0.0575 & 0.0904 & 0.0375 & 0.0480
& 0.0308 & 0.0482 & 0.0197 & 0.0255 \\

+ Gap Tokens
& 0.0413 & 0.0643 & 0.0270 & 0.0344 
& 0.0609 & 0.0945 & \textbf{0.0400} & 0.0507
& 0.0317 & 0.0499 & 0.0204 & 0.0263 \\

ChronoSID              
& \textbf{0.0419} & \textbf{0.0651} & \textbf{0.0273} & \textbf{0.0347} 
& \textbf{0.0609} & \textbf{0.0952} & 0.0399 & \textbf{0.0509}
& \textbf{0.0340} & \textbf{0.0512} & \textbf{0.0218} & \textbf{0.0273} \\
\bottomrule
\end{tabular}
\end{table*}

\subsection{Main Results (RQ1)}
\label{sec:main_results}

Table~\ref{tab:main_results} reports the overall recommendation performance on eight Amazon product domains.
ChronoSID achieves the best performance among SID-based generative recommendation methods across all reported datasets and metrics.
Compared with conventional sequential recommenders, ChronoSID also obtains the best or near-best results on most metrics, showing that the semantic-ID generation paradigm remains competitive with strong item-ID-based sequence models.

We first compare ChronoSID with conventional sequential recommendation baselines.
Standard sequential recommenders, including HGN, SASRec, BERT4Rec, and S$^3$-Rec, generally underperform SID-based generative methods.
This suggests that representing items with compact semantic ID tuples and generating target identifiers can be effective for sequential recommendation.
Some strengthened sequential variants remain competitive on a few metrics, such as R@10 on IS and TG, indicating that strong item-level sequence encoders can still be effective in certain domains.
Nevertheless, ChronoSID achieves the best or near-best performance in most cases while preserving the compact generative recommendation paradigm.

The comparison with generative baselines is more central to our study.
ChronoSID consistently outperforms TIGER, LETTER, EAGER, UNGER, ETEGRec, and ReSID across the eight domains.
In particular, ReSID is the most direct baseline because it shares the same three-stage semantic ID pipeline and the same GAOQ quantization stage.
Therefore, the improvements over ReSID directly reflect the benefit of introducing temporal information into representation learning and sequence generation.
For example, on MI, ChronoSID improves ReSID from 0.0388 to 0.0417 in R@5 and from 0.0325 to 0.0346 in N@10.
On VG, ChronoSID improves R@10 from 0.0898 to 0.0927 and N@10 from 0.0480 to 0.0501.
Similar gains are observed on IS, BP, ACS, SO, TG, and BPC, although the absolute improvement varies across domains.

Overall, the main results answer RQ1 positively.
ChronoSID improves over strong SID-based generative baselines while keeping the underlying semantic ID quantization stage unchanged.
These results support our central hypothesis that static semantic ID sequences are temporally incomplete, and that augmenting them with historical gap tokens provides useful behavioral cues for generative recommendation.

\subsection{Ablation Study (RQ2)}
\label{sec:ablation}

Table~\ref{tab:ablation} studies the contribution of the two temporal components in ChronoSID on MI, VG, and IS. 
Starting from ReSID, \textbf{+ TA-FAMAE} adds the temporal auxiliary objective during item representation learning while keeping the original ReSID-style sequence generator. 
\textbf{+ Gap Tokens} keeps the original ReSID item representation learning stage but injects discretized inter-interaction gap tokens into the T5 encoder input. 
The full \textbf{ChronoSID} combines both components.

The results show that both temporal components are beneficial, but they contribute differently. 
Adding TA-FAMAE improves over ReSID on most metrics, indicating that time-gap prediction provides useful regularization for item representation learning. 
However, the gains from TA-FAMAE alone are relatively moderate, which is expected because the temporal signal is introduced only indirectly through the item representation stage. 
By contrast, gap-token injection yields larger improvements across the three datasets. 
For example, on VG, \textbf{+ Gap Tokens} improves R@10 from 0.0898 to 0.0945 and N@10 from 0.0480 to 0.0507. 
This confirms that explicitly exposing historical inter-interaction gaps to the sequence-to-sequence generator is the main source of improvement.

The full ChronoSID generally achieves the best overall performance. 
On MI, ChronoSID improves ReSID from 0.0388 to 0.0419 in R@5 and from 0.0325 to 0.0347 in N@10. 
On IS, the improvement is also consistent across all metrics, with R@5 increasing from 0.0305 to 0.0340 and N@10 from 0.0250 to 0.0273. 
These results suggest that the two temporal components are complementary: TA-FAMAE provides representation-level temporal regularization, while gap-token injection directly supplies temporal context to the generator.

\subsection{Temporal Robustness Analysis (RQ3)}
\label{sec:long_gap}

\begin{figure*}[t]
    \centering
    \includegraphics[width=\linewidth]{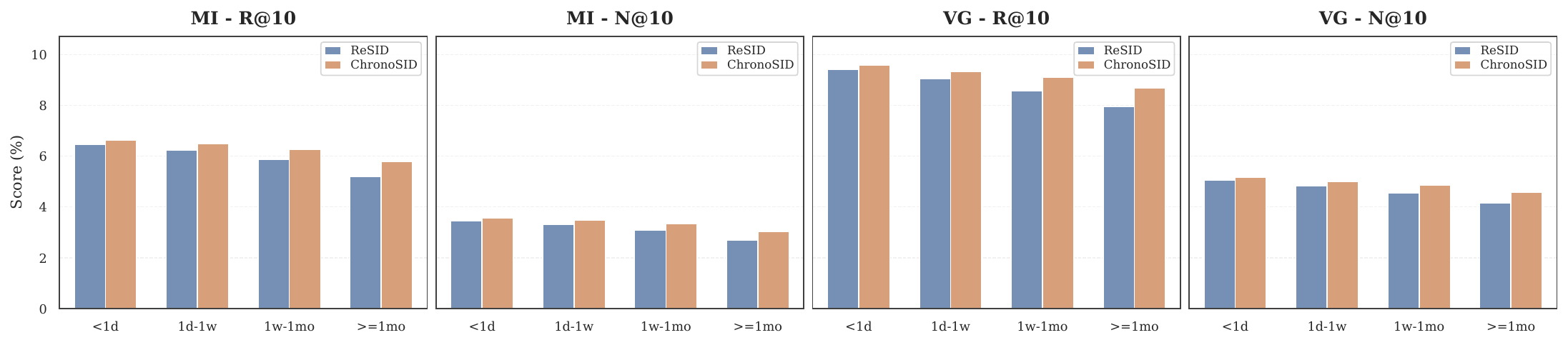}
    \vspace{-1.5em}
    \caption{
    Temporal robustness analysis on MI and VG.
    Test cases are grouped by the elapsed time between the last historical interaction and the target interaction.
    We report R@10 and N@10 under different gap groups.
    The target-side interval is used only for diagnostic grouping and is not provided as model input.
    }
    \label{fig:long_gap}
    \vspace{-0.8em}
\end{figure*}

To examine whether temporal modeling improves robustness under different temporal conditions, we group test instances according to the elapsed time between the last historical interaction and the target interaction.
This target-side interval is used only for diagnostic grouping and is not provided as an additional input feature during evaluation.
Longer intervals usually indicate weaker short-term continuity and a higher likelihood of preference drift, making these cases more challenging for models that mainly rely on local item transitions.

Figure~\ref{fig:long_gap} reports the results on MI and VG using R@10 and N@10.
ChronoSID consistently outperforms ReSID across all gap groups on both datasets, showing that temporal gap modeling improves recommendation quality under different temporal conditions.
As the gap becomes longer, the absolute performance of both models generally decreases, which confirms that delayed-return cases are more difficult than short-gap cases.
Nevertheless, ChronoSID maintains clear gains over ReSID in the long-gap groups, especially for $[1w,1mo)$ and $\geq 1mo$ interactions.

These observations support our motivation that historical inter-interaction gaps provide useful behavioral cues for modeling interest continuity and drift.
Rather than treating a user history as a purely static sequence of item codes, ChronoSID augments the semantic ID sequence with temporal gap tokens, allowing the generator to exploit temporal patterns observed in the history.
The stronger gains under delayed-return cases suggest that such temporal augmentation is particularly helpful when local item transitions become less reliable.

\subsection{Sensitivity Analysis (RQ4)}
\label{sec:sensitivity}

\begin{table}[t]
\centering
\caption{
Sensitivity analysis of ChronoSID on MI and VG datasets. 
We vary the temporal auxiliary loss weight $\lambda$ and the number of gap bins $G$. 
The best result in each block is shown in bold.
}
\label{tab:sensitivity}
\resizebox{\linewidth}{!}{
\begin{tabular}{l c cccc cccc}
\toprule
\multirow{2}{*}{Setting} & \multirow{2}{*}{Value}
& \multicolumn{4}{c}{MI}
& \multicolumn{4}{c}{VG} \\
\cmidrule(lr){3-6} \cmidrule(lr){7-10}
& & R@5 & R@10 & N@5 & N@10 & R@5 & R@10 & N@5 & N@10 \\
\midrule
\multirow{5}{*}{$\lambda$}
& 0     & 0.0413 & 0.0643 & 0.0270 & 0.0344 & 0.0609 & 0.0945 & 0.0400 & 0.0507 \\
& 0.01  & 0.0415 & 0.0646 & 0.0271 & 0.0345 & 0.0608 & 0.0948 & 0.0399 & 0.0508 \\
& 0.05  & 0.0417 & 0.0649 & 0.0272 & 0.0346 & 0.0610 & 0.0950 & 0.0401 & 0.0509 \\
& 0.1   & \textbf{0.0419} & \textbf{0.0651} & \textbf{0.0273} & \textbf{0.0347} & 0.0609 & \textbf{0.0952} & 0.0399 & \textbf{0.0509} \\
& 0.2   & 0.0414 & 0.0645 & 0.0270 & 0.0344 & 0.0602 & 0.0941 & 0.0395 & 0.0503 \\
\midrule
\multirow{4}{*}{$G$}
& 3  & 0.0412 & 0.0640 & 0.0269 & 0.0342 & 0.0601 & 0.0938 & 0.0394 & 0.0501 \\
& 5  & \textbf{0.0419} & \textbf{0.0651} & \textbf{0.0273} & \textbf{0.0347} & \textbf{0.0609} & \textbf{0.0952} & \textbf{0.0399} & \textbf{0.0509} \\
& 7  & 0.0416 & 0.0647 & 0.0271 & 0.0345 & 0.0607 & 0.0947 & 0.0398 & 0.0506 \\
& 10 & 0.0411 & 0.0642 & 0.0268 & 0.0341 & 0.0600 & 0.0939 & 0.0392 & 0.0499 \\
\bottomrule
\end{tabular}
}
\end{table}

Table~\ref{tab:sensitivity} analyzes the sensitivity of ChronoSID to two temporal hyperparameters: the temporal auxiliary loss weight $\lambda$ in TA-FAMAE and the number of discretized gap bins $G$ used for gap-token injection. For the temporal loss weight $\lambda$, setting $\lambda=0$ removes the auxiliary time-gap prediction objective from TA-FAMAE while keeping gap-token injection unchanged.
Compared with this variant, using a positive $\lambda$ generally improves performance on MI and maintains competitive performance on VG, suggesting that temporal prediction provides useful regularization for item representation learning.
However, the effect is moderate, which is consistent with the ablation study: the main performance gain comes from directly injecting gap tokens into the sequence generator, while TA-FAMAE provides additional representation-level benefits.
When $\lambda$ becomes too large, performance slightly drops, indicating that over-emphasizing temporal prediction may distract the representation learner from preserving item semantics.
Overall, $\lambda=0.1$ provides a stable trade-off and is used as the default setting.

For the number of gap bins $G$, ChronoSID performs best with a moderate temporal granularity.
Using too few bins, such as $G=3$, may merge behaviorally different intervals and weaken the temporal signal.
In contrast, using too many bins, such as $G=10$, can make temporal tokens sparse and harder to learn reliably.
The setting $G=5$ achieves the best overall performance on both datasets, which supports our design choice of using fixed log-scale temporal bins with thresholds $\{1\mathrm{h},1\mathrm{d},1\mathrm{w},1\mathrm{mo}\}$.
This result suggests that coarse but interpretable temporal discretization is sufficient to capture useful gap patterns for SID-based generative recommendation.

\subsection{Efficiency Analysis (RQ5)}
\label{sec:efficiency}

Finally, we analyze the computational overhead introduced by ChronoSID compared with ReSID.
ChronoSID inserts one additional gap token before each item semantic ID tuple, increasing the encoder-side input length from $3L$ to $4L$ tokens.
However, the decoder target remains unchanged because both models generate the same three-level semantic ID tuple.
Therefore, the additional cost mainly comes from encoder-side computation and the temporal auxiliary objective, while the decoder-side generation target and beam search depth remain the same. All efficiency measurements are conducted on the same hardware with an identical batch size, beam size, and evaluation protocol.

\begin{table}[t]
\centering
\caption{
Efficiency comparison between ReSID and ChronoSID on MI and VG.
Training time is measured in minutes.
Inference time denotes the total generation time on the evaluation split, and the per-sample latency is reported in milliseconds.
Ratios are computed relative to ReSID on the same dataset.
}
\vspace{-1em}
\label{tab:cost}
\resizebox{\linewidth}{!}{
\begin{tabular}{l l c c c c c c}
\toprule
Dataset & Model 
& Train Time
& Train Ratio
& Infer. Time
& ms/sample
& R@10 
& N@10 \\
\midrule
\multirow{2}{*}{MI}
& ReSID     
& 174.0 min 
& 1.00$\times$ 
& 39.61 s 
& 0.69 
& 0.0614 
& 0.0325 \\
& ChronoSID 
& 187.0 min 
& 1.07$\times$ 
& 49.13 s 
& 0.86 
& \textbf{0.0651} 
& \textbf{0.0347} \\
\midrule
\multirow{2}{*}{VG}
& ReSID     
& 275.0 min 
& 1.00$\times$ 
& 156.39 s 
& 1.68 
& 0.0898 
& 0.0480 \\
& ChronoSID 
& 356.0 min 
& 1.29$\times$ 
& 199.51 s 
& 2.15 
& \textbf{0.0952} 
& \textbf{0.0509} \\
\bottomrule
\end{tabular}
}
\end{table}

Table~\ref{tab:cost} reports the efficiency comparison on MI and VG.
On MI, ChronoSID increases training time from 174.0 minutes to 187.0 minutes, corresponding to a $1.07\times$ training overhead.
Its total inference time increases from 39.61 seconds to 49.13 seconds, with per-sample latency increasing from 0.69 ms to 0.86 ms.
On VG, ChronoSID increases training time from 275.0 minutes to 356.0 minutes, corresponding to a $1.29\times$ overhead.
For the generation on the VG evaluation split, ChronoSID increases the total inference time from 156.39 seconds to 199.51 seconds, with per-sample latency increasing from 1.68 ms to 2.15 ms.

Overall, ChronoSID introduces additional computational cost due to temporal auxiliary learning and longer encoder inputs.
Nevertheless, the overhead remains moderate because the decoder still generates the same short three-level semantic ID tuple and uses the same beam search depth as ReSID.
Considering the accuracy improvements in R@10 and N@10, ChronoSID provides temporal awareness with acceptable computational overhead while preserving the compact generative recommendation paradigm.

\subsection{Performance over Popular and Unpopular Items (RQ6)}
\label{sec:popularity}

\begin{figure}[t]
    \centering
    \includegraphics[width=\linewidth]{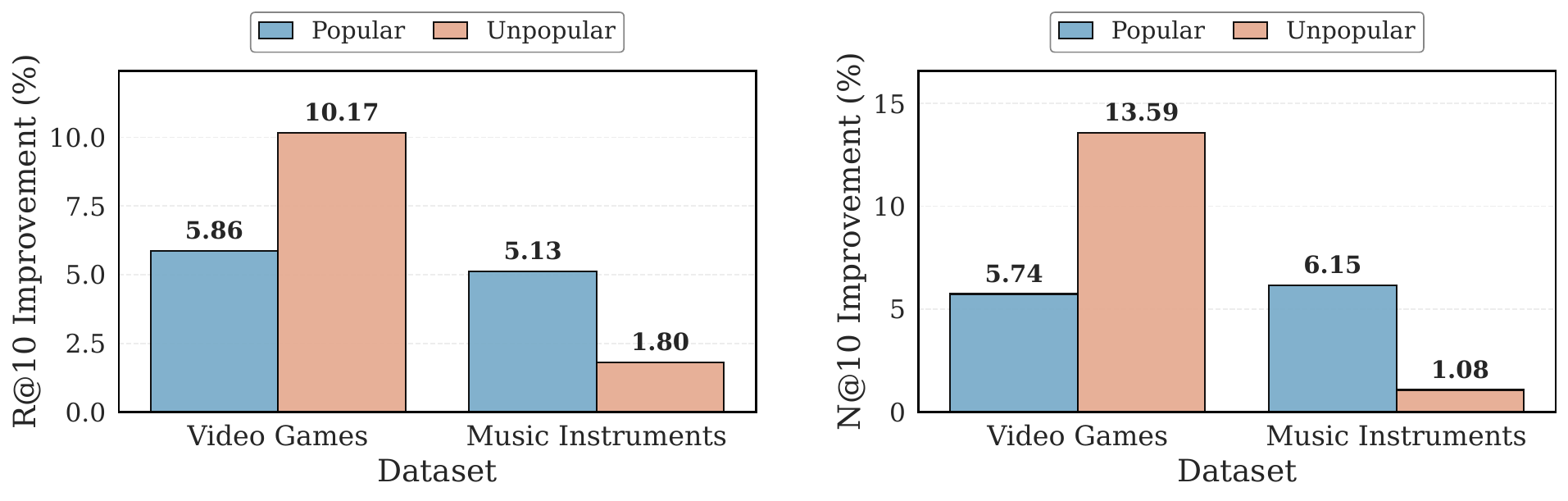}
    \vspace{-2em}
    \caption{
    Relative improvement of ChronoSID over ReSID on popular and unpopular target items.
    Improvements are reported for R@10 and N@10 on Video Games and Musical Instruments.
    }
    \label{fig:popularity}
\end{figure}

We further analyze whether the gains of ChronoSID are mainly driven by popular items or whether temporal gap modeling also benefits less frequent items.
This analysis helps examine whether ChronoSID simply strengthens popularity-driven recommendations or provides useful signals across different item-frequency regimes.
Following common practice, we define the top 20\% most frequently interacted items in the training set as popular items and treat the remaining 80\% as unpopular items.
We then group test instances according to whether their target item belongs to the popular or unpopular set and compute the relative improvement of ChronoSID over ReSID within each group.

Figure~\ref{fig:popularity} reports the results on Video Games and Musical Instruments.
ChronoSID improves over ReSID for both popular and unpopular target items, suggesting that temporal augmentation is not merely amplifying popularity bias.
On Video Games, the gains are larger for unpopular items: ChronoSID improves R@10 by $10.17\%$ and N@10 by $13.59\%$ on unpopular items, compared with $5.86\%$ and $5.74\%$ on popular items.
This suggests that temporal gap tokens provide useful behavioral context when item-frequency signals are less reliable.

On Musical Instruments, ChronoSID also improves both item groups, but the gains are larger for popular items.
This domain-dependent pattern indicates that the effect of temporal modeling varies with the interaction distribution and popularity structure of each domain.
Overall, the results answer RQ6 positively: ChronoSID improves recommendation quality in both popularity regimes, showing that temporal gap modeling provides complementary behavioral signals beyond a single popularity-driven pattern.

\subsection{Output-Level Analysis (RQ7)}
\label{sec:output_analysis}

To further examine how ChronoSID improves SID-based generative recommendation, we conduct an output-level analysis on the generated semantic ID candidates.
This analysis compares ChronoSID and ReSID on the same test instances under the same decoding and semantic ID lookup protocol.
Its goal is to verify whether temporal augmentation helps the generator produce more accurate target SID tuples, rather than only improving aggregate ranking metrics.

Table~\ref{tab:output_level_analysis} reports the output-level comparison.
ChronoSID improves over ReSID on all four metrics, indicating that temporal gap modeling helps the generator produce more accurate semantic ID candidates.
Since both methods are evaluated on identical test instances and use the same semantic ID lookup table, the improvement can be attributed to the temporally augmented input sequence rather than differences in decoding or candidate resolution.
Moreover, the paired bootstrap confidence intervals are strictly positive across all metrics, suggesting that the observed gains are stable rather than caused by random evaluation fluctuations.

Overall, the output-level analysis answers RQ7 positively.
By interleaving historical gap tokens with semantic ID sequences, ChronoSID improves the accuracy of generated SID candidates under the same generation and lookup setting as ReSID.
This provides additional evidence that temporal augmentation directly benefits the generative prediction process.

\begin{table}[t]
\centering
\vspace{-1em}
\caption{
Output-level comparison between ReSID and ChronoSID on the MI dataset.
The confidence interval is computed for the absolute improvement of ChronoSID over ReSID using paired bootstrap resampling.
}
\label{tab:output_level_analysis}
\vspace{-0.8em}
\small
\setlength{\tabcolsep}{5pt}
\begin{tabular}{lcccc}
\toprule
\textbf{Metric} & \textbf{ReSID} & \textbf{ChronoSID} & \textbf{Abs. Gain} & \textbf{95\% CI} \\
\midrule
R@5  & 0.0386 & 0.0413 & +0.0027 & [0.0013, 0.0041] \\
R@10 & 0.0612 & 0.0642 & +0.0031 & [0.0014, 0.0046] \\
N@5  & 0.0253 & 0.0271 & +0.0019 & [0.0010, 0.0028] \\
N@10 & 0.0325 & 0.0345 & +0.0020 & [0.0011, 0.0028] \\
\bottomrule
\end{tabular}
\vspace{-1em}
\end{table}

\section{Conclusion}

In this paper, we propose ChronoSID, a lightweight temporal augmentation framework for semantic-ID-based generative recommendation. 
ChronoSID addresses the temporal blindness of existing SID-based methods by incorporating inter-interaction time gaps into both item representation learning and sequence generation. 
Specifically, TA-FAMAE regularizes item representations with an auxiliary time-gap prediction objective, while gap-token interleaving injects discretized historical time intervals into the T5 encoder input. 
Experiments on Amazon review benchmarks show that ChronoSID improves over strong generative recommendation baselines, especially ReSID. 
Ablation, long-gap, and diagnostic analyses further demonstrate that gap-token injection is the main source of improvement and that temporal modeling is particularly helpful when user interests are more likely to drift. 
Future work includes adaptive temporal discretization and joint modeling of next-item and next-time prediction.

\bibliographystyle{ACM-Reference-Format}
\bibliography{sample-base}


\end{document}